\documentclass[a4paper, amsfonts, amssymb, amsmath, reprint, showkeys, nofootinbib, twoside]{revtex4-1}
\usepackage[english]{babel}
\usepackage[utf8]{inputenc}
\usepackage[colorinlistoftodos, color=green!40, prependcaption]{todonotes}
\usepackage{amsthm}
\usepackage{mathtools}
\usepackage{physics}
\usepackage{xcolor}
\usepackage{graphicx}
\usepackage[left=23mm,right=13mm,top=35mm,columnsep=15pt]{geometry} 
\usepackage{adjustbox}
\usepackage{placeins}
\usepackage[T1]{fontenc}
\usepackage{lipsum}
\usepackage{csquotes}

\usepackage[pdftex, pdftitle={Article}, pdfauthor={Author}]{hyperref} % For hyperlinks in the PDF
\begin{document}
\title{TOF-SIMS Analysis of Decoherence Sources in Nb Superconducting Resonators}

\author{Akshay A. Murthy$^1$}
\email[Correspondence email address: ]{amurthy@fnal.gov}
\author{Jae-Yel Lee$^1$, Cameron Kopas$^2$, Matthew J. Reagor$^2$, Anthony P. McFadden$^{3}$, David P. Pappas$^3$, Mattia Checchin$^1$, Anna Grassellino$^1$, Alexander Romanenko$^1$}

\affiliation{$^1$Fermi National Accelerator Laboratory (FNAL), Batavia, IL 60510, USA}
\affiliation{$^2$Rigetti Computing, Berkeley, CA 94710, USA}
\affiliation{$^3$National Institute of Standards and Technology, Boulder, CO 80305, USA}

\date{\today} % Leave empty to omit a date

\begin{abstract}
Superconducting qubits have emerged as a potentially foundational platform technology for addressing complex computational problems deemed intractable with classical computing. Despite recent advances enabling multiqubit designs that exhibit coherence lifetimes on the order of hundreds of $\mu$s, material quality and interfacial structures continue to curb device performance. When niobium is deployed as the superconducting material, two-level system defects in the thin film and adjacent dielectric regions introduce stochastic noise and dissipate electromagnetic energy at the cryogenic operating temperatures. In this study, we utilize time-of-flight secondary ion mass spectrometry (TOF-SIMS) to understand the role specific fabrication procedures play in introducing such dissipation mechanisms in these complex systems. We interrogated Nb thin films and transmon qubit structures fabricated by Rigetti Computing and at the National Institute of Standards and Technology through slight variations in the processing and vacuum conditions. We find that when Nb film is sputtered onto the Si substrate, oxide and silicide regions are generated at various interfaces. We also observe that impurity species such as niobium hydrides and carbides are incorporated within the niobium layer during the subsequent lithographic patterning steps. The formation of these resistive compounds likely impact the superconducting properties of the Nb thin film. Additionally, we observe the presence of halogen species distributed throughout the patterned thin films. We conclude by hypothesizing the source of such impurities in these structures in an effort to intelligently fabricate superconducting qubits and extend coherence times moving forward.
\end{abstract}


\maketitle

\section{Introduction} \label{sec:intro}
Superconducting qubits represent one of the most mature platforms of emerging quantum information technologies.\cite{doi:10.1146/annurev-conmatphys-031119-050605, Wendin_2017, deLeoneabb2823} This platform has enabled computers that can outperform the world's largest classical computers with respect to calculating random circuits\cite{RN5} as well as processors consisting of 65 superconducting qubits with gate and readout fidelities of 99.97\% and 99.8\%, respectively.\cite{Jurcevic_2021} Nonetheless, the presence of surfaces, interfaces, and defects in the constituent materials have proven to curtail coherence times well below the microsecond and second timescales necessary for scalable quantum devices.\cite{RN6,PhysRevLett.123.190502, 8993458} In particular, these structures introduce two-level systems, which contribute heavily to microwave loss at the cryogenic operating temperatures and the single photon level operating powers.\cite{Muller_2019} As such, this has accelerated worldwide efforts to identify and eliminate these sources of decoherence in superconducting qubit systems.\cite{5685585, doi:10.1063/1.4934486, Dial_2016, PhysRevApplied.12.014012}

In terms of materials, Nb thin films have been employed extensively as the superconducting component in such qubit architectures. These films can be deposited via a variety of techniques including high power impulse sputtering,\cite{doi:10.1063/1.2697052} electron cyclotron resonance,\cite{WU200556} and cathodic arcs.\cite{Krishnan_2011} Their low kinetic impedance \cite{Annunziata_2010} and compatibility with industrial level processes particularly when deposited on silicon wafers makes them quite attractive for superconducting qubit applications.\cite{6967749} Furthermore, the considerable development in niobium metal refining over the past few decades has enabled the extraction of ultra high purity niobium sources at the fraction of the cost associated with alternative superconducting metals such as Mo, Re, and Ta.\cite{ZHU20111} Nevertheless, the experimentally derived microwave loss associated with Nb thin films is typically orders of magnitude larger than that expected for the pure elemental compound. This has generally been attributed to the presence of native amorphous oxides with varying stoichiometries (NbO, NbO$_2$, Nb$_2$O$_5$) that are known to host two-level system defects.\cite{Niepce_2020, Burnett_2016, premkumar2020microscopic, verjauw2020investigation} Although comparatively underinvestigated, it is likely that additional chemical impurities in these films that impact the superconducting properties and help explain the measured loss in these films.\cite{Romanenko_2013, lee2021discovery} As such, a systematic investigation of the chemical constituents present within these thin films combined with an identification of processing protocols that give rise to these defect structures is critical for enabling more intelligent design of superconducting qubit architectures.

To this end, we employ time-of-flight secondary ion mass spectrometry (TOF-SIMS) to thoroughly understand the chemical constituents present within Nb thin films deposited on Si wafers through sputter deposition. By performing this detailed characterization on samples taken from various steps throughout the superconducting process, we are able to readily identify the role that fabrication protocols play in terms of introducing structural deviations and contamination in such systems. These samples include a blanket film of Nb sputter deposited on a Si substrate, along with samples which have undergone lithographic patterning steps and a fully fabricated superconducting qubit architecture. We find that oxide and silicide regions are generated at the metal/air and metal/substrate interfaces following deposition of Nb. We also find that the subsequent lithographic patterning steps lead to an inward diffusion of hydrogen and carbon throughout the thin film, which generates niobium hydrides and carbides. Additionally, we observe that the etching procedures lead to a similar inward diffusion of halogen species throughout the patterned thin films, which thereby generates a final architecture composed of a multitude of sources potentially leading to quantum decoherence. 
\section{Experimental} \label{sec:experimental}
Nb films were prepared on the Si (001) wafers (Vendor) via HiPIMS with a vacuum level of 1E-8 Torr as well as sputter deposition.  Prior to deposition, the wafer was prepared with an RCA surface treatment detailed previously.\cite{doi:10.1063/1.3517252, 8993458} An annular dark field (ADF) scanning/transmission electron microscopy images of this interface is provided in Figure \ref{fgr:STEM_SIMS}a-c. Following this step, resonator and capacitance pads were fabricated following the procedures detailed by \citeauthor{8993458} \cite{8993458} SIMS measurements were performed after each step using a dual beam time-of-flight secondary ion mass spectrometry (IONTOF 5) to analyze the concentration and depth distribution of impurities in the Nb film. Secondary ion measurements were performed using a liquid bismuth ion beam (Bi+). A cesium ion gun with an energy of 500 eV was used for sputtering the surface for depth profile measurements to detect anions and an oxygen ion gun with an energy of 500 eV was used for sputtering the surface for depth profile measurements to enhance detection of metallic impurities.\cite{Bose_2015}
\section{Results} \label{sec:results}
% \subsection{Film Surface}
% \subsection{Film/Substrate Interface}
% \subsection{Film Contamination}

\begin{figure*}
\includegraphics[width=7in]{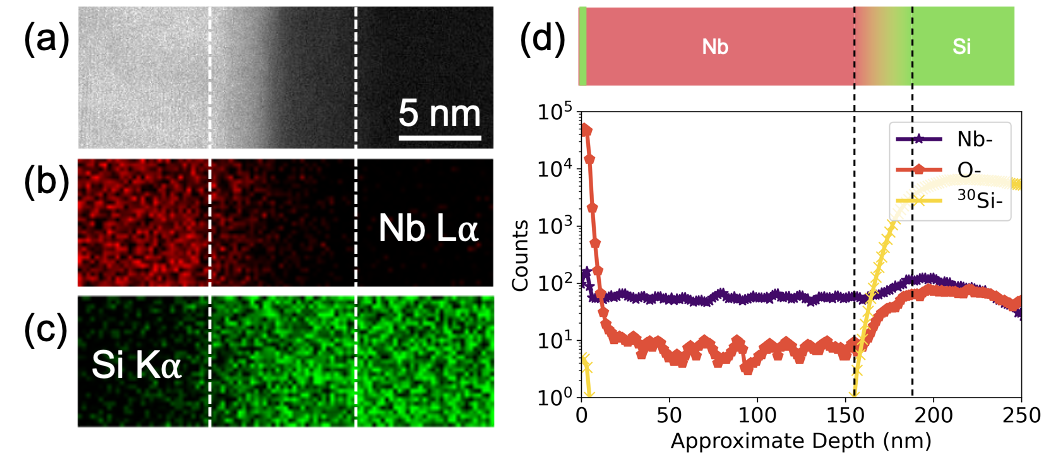}
\caption{Nb/Si interface (a) STEM image taken at the film/substrate interface of Nb thin film deposited on a Si (001) wafer. (b) and (c) Associated EDS maps constructed from this interface using characteristic Nb L$\alpha$ and Si K$\alpha$ x-ray emission. (d) TOF-SIMS depth profile taken from this interface. Alloyed region is indicated by dotted lines.}
\label{fgr:STEM_SIMS}
\end{figure*}

TOF-SIMS measurements were taken immediately following Nb deposition and a depth profile of the film is provided in Figure \ref{fgr:STEM_SIMS}d. In this figure, the Nb film region and Si substrate are clearly indicated. One interesting feature from the TOF-SIMS depth profile measurements, is that it appears that Si atoms float to the very surface of the Nb film and oxidize to form SiO$_2$. The mass spectrum taken from the top 5nm of the surface compared to that taken from within the film is provided in Figure S1. Although the observation of this phenomenon has not previously been reported in literature associated with superconducting qubits, the propensity for Si to readily diffuse during film deposition has been frequently observed in other systems. For instance, there have been numerous reports of adsorbed Si atoms diffusing through and intercalating between thin films and substrates.\cite{C8NR00648B, doi:10.1063/1.3687190, RN7} We suspect that a similar process is at play during this deposition. In particular, we hypothesize that during Nb deposition, a few layers of Si atoms continuously float to the top surface in an effort to reduce the surface free energy of the system. While this feature may go undetected in other surface analysis techniques, this is likely identifiable with TOF-SIMS due to the parts-per-million range trace element detection capability that this technique offers.\cite{Bose_2015} As the Si atoms at the surface are quite susceptible to oxidation which can serve as an additional loss mechanism, this feature requires further investigation with more sensitive and quantitative techniques including atomic probe tomography and Rutherford backscattering spectroscopy.\cite{RN8}

Additionally, from the TOF-SIMS data, we observe a gradual interface between Nb and Si where Nb atoms are implanted within the Si substrate. Although collisional cascade effects including atomic mixing during TOF-SIMS depth profiling can preclude direct quantification of this graded interface,\cite{doi:10.1116/1.586352} STEM EDS results corroborate the presence of an alloyed interface on the order of 8 nm. Based on previous findings, it is likely that this region is composed of alloys such as the Nb$_5$Si$_3$ and NbSi$_2$ phases.\cite{PRASAD20111577} As the signal intensity in ADF images is proportional to Z$^{\alpha}$, where Z represents atomic number and $\alpha$ lies between 1.2 and 1.8,\cite{RIBET2021} the presence of this alloyed region is supported by Figure \ref{fgr:STEM_SIMS}a as well. It is possible that this alloyed region could appreciably reduce the superconducting transition temperature or serve as an additional scattering source leading to the breaking of Cooper pairs.\cite{PhysRevB.45.535}

\begin{figure*}
\includegraphics[width=7in]{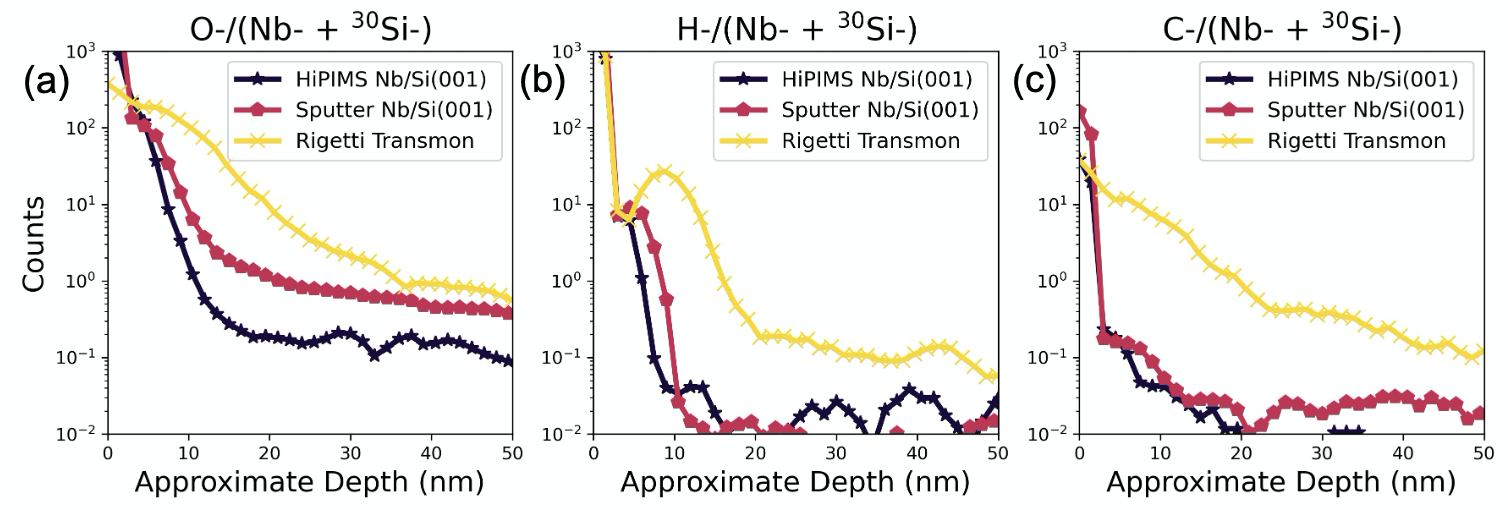}
\caption{TOF-SIMS depth profiles of hydrocarbon species in Nb thin films. (a) Oxygen, (b) Hydrogen, and (c) Carbon signal are plotted as a function of depth for the three samples investigated. These include: Nb thin film deposited with HiPIMS, Nb thin film deposited with sputtering, and Nb capacitance pad from patterned transmon qubit.}
\label{fgr:OHC}
\end{figure*}

In addition to Si at the surface and the interfacial amorphous alloys, we examined the distribution of oxides and hydrocarbons in the blanket Nb film. Both of these species are known to introduce loss in superconducting niobium systems and are presented in Figure \ref{fgr:OHC}.\cite{Niepce_2020, Burnett_2016, premkumar2020microscopic, verjauw2020investigation,PhysRevApplied.13.034032,PhysRevLett.119.264801} From this figure, we observe an appreciable content of oxygen in the immediate vicinity of the surface. This can be attributed to previous detailed investigations detailing the presence of various Nb oxides in this region.\cite{Niepce_2020, Burnett_2016, premkumar2020microscopic, verjauw2020investigation} Further, it is likely that the presence of Si atoms at the surface and the SiO$_x$ that can form contribute to this oxide intensity as well. Moreover, as the propensity for Nb to form hydrides and carbides has been studied extensively,\cite{doi:10.1063/1.4918272, doi:10.1063/1.4826901, lee2021discovery} it is no surprise to observe enhanced signals corresponding to carbon and hydrogen at the surface. 

Similar measurements taken from Nb deposited via sputter deposition are provided in Figure \ref{fgr:OHC} as well. In this case, we observe similar H and C profiles to that observed for Nb films deposited with HiPIMS. The oxygen counts, however, are roughly one order of magnitude higher in the Nb films deposited with sputter deposition. This can likely be attributed to differences in vacuum conditions. Based on previous findings, it is likely that this oxygen is mainly localized at grain boundaries in the Nb film.\cite{RN11}

\begin{figure*}
\includegraphics[width=7in]{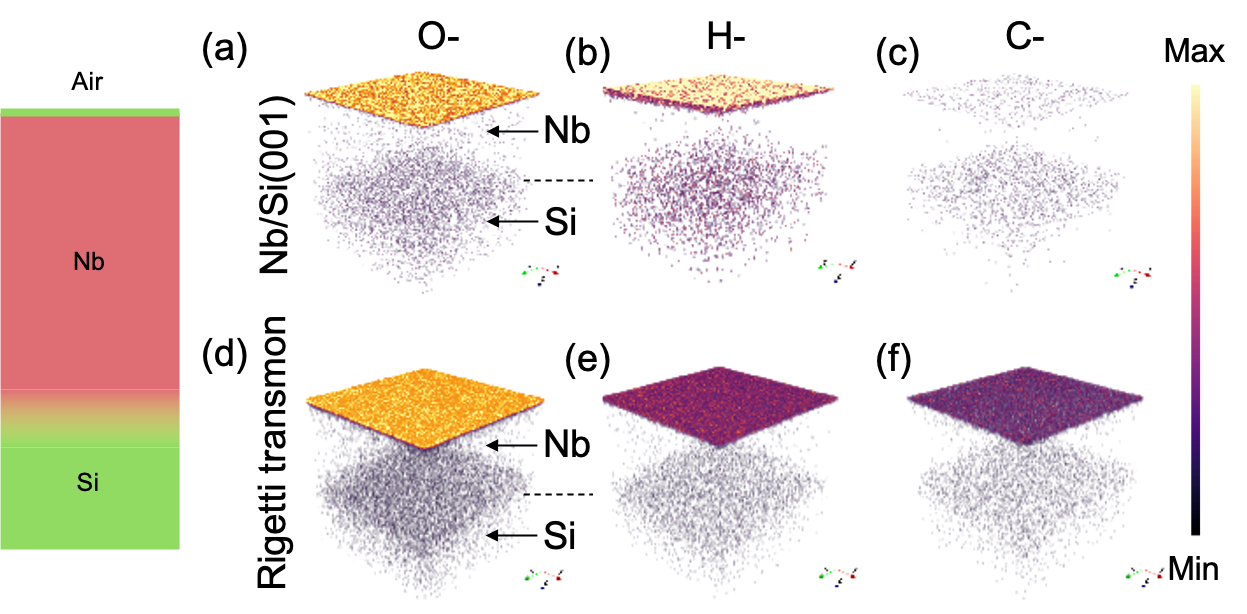}
\caption{3D rendering representing TOF-SIMS depth profiles of hydrocarbon species in Nb thin films. (a) Oxygen, (b) Hydrogen, and (c) Carbon signal are plotted as a function of depth for the Nb thin film deposited with HiPIMS. (d) Oxygen, (e) Hydrogen, and (f) Carbon signal are plotted as a function of depth for the Nb capacitance pad from patterned transmon qubit.}
\label{fgr:3D-OHC}
\end{figure*}

Following the lithographic patterning of the qubit architecture, TOF-SIMS depth profiles are taken from the Nb capacitance pads deposited with HiPIMS to once again probe the distribution of oxides and hydrocarbons in the film. In this case, we find that the concentration of hydrogen, carbon, and oxygen increase uniformly. Specifically, we find that in addition to observing a large intensity of these species at the surface, this signal associated with these species decays much more gradually before asymptotically approaching a baseline value. In all three cases, this decay length extends through the upper 30nm of the Nb film. Additionally, the baseline signal of these species is roughly one order of magnitude larger than the respective signals measured in the blanket Nb film. Three-dimensional images representing the distribution of these species are provided in Figure \ref{fgr:3D-OHC}. The relative in-plane uniformity present in these profiles confirms that depth profiles are taken from representative regions free of surface features and the artifacts that they can introduce.\cite{BOSE2020145464,romanenko:srf2019-thp014} Furthermore, it is again apparent that the lithography process introduces increased oxygen and hydrocarbon species throughout the film.

Based on the fabrication process, we hypothesize that these contaminant species are likely introduced through the organic photoresist used during the lithography process. We suspect that photoresist residue may remain on the thin film surface following lift-off procedures and can diffuse into the film during subsequent baking steps. One potential solution to reduce the presence of these species would be to define all relevant patterns using a hard, physical masks.\cite{bilmes2021insitu,doi:10.1063/5.0037093}

In addition to the aforementioned contaminants, we also probe the distribution of other anionic and cationic contaminants in the Nb film. In the case of hydroxides, fluorides and chlorides, we observe a very similar behavior in the blanket films compared to the lithographically defined capacitance pads. Specifically, we observe that whereas all samples exhibit >10$^1$ counts of OH-, F- and Cl- at the surface, in the blanket film sample, the intensity of these contaminants decays to a stable baseline value within a few nanometers, whereas a much more gradual decay over a period of 30 nm is observed in the lithographically defined patterns. Additionally, the baseline value of these signals is an order of magnitude larger in the patterned films compared to the blanket films and potentially impact the conductivity of the superconducting film. We hypothesize that these halogen species are introduced through the reactive ion etching and wet etching processes, which further highlights the value in potentially employing physical masks to define relevant features in the qubit structure.\cite{bilmes2021insitu,doi:10.1063/5.0037093,kopas2020characterization}

\begin{figure*}
\includegraphics[width=7in]{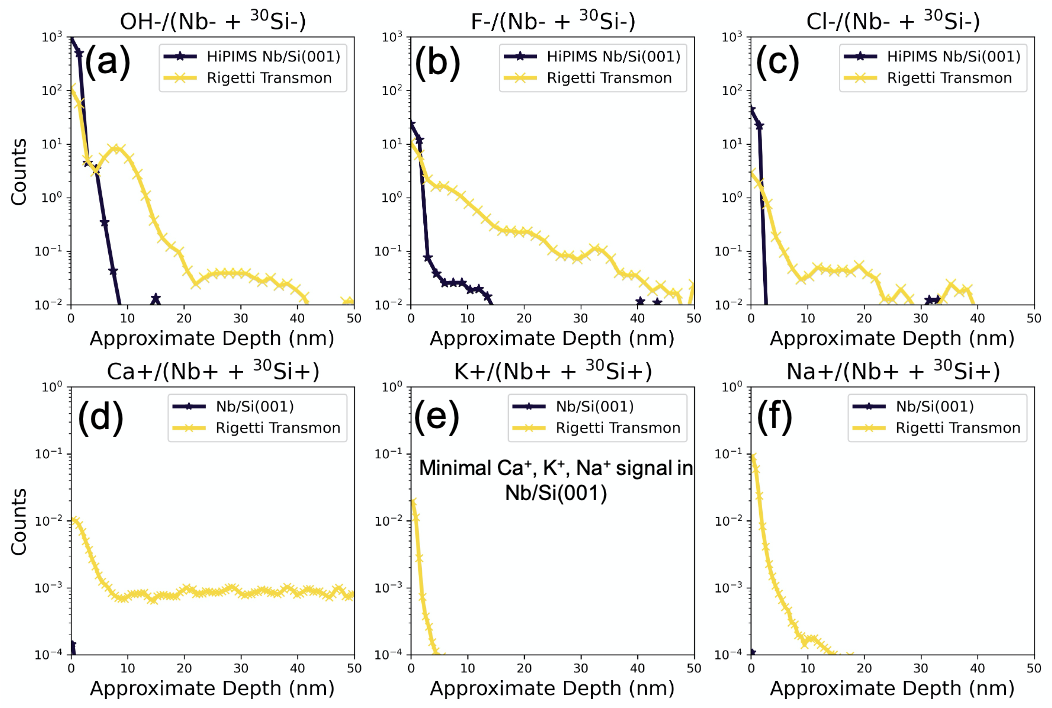}
\caption{TOF-SIMS depth profiles of various anionic and cationic species in Nb thin films. (a) OH-, (b) F-, (c) Cl-, (d) Ca+, (e) K+, and (f) Na+ signals are plotted as a function of depth for Nb thin film deposited with HiPIMS and Nb capacitance pad from patterned transmon qubit.}
\label{fgr:contaminant}
\end{figure*}

Finally, a distribution of cationic contaminants are provided in Figure \ref{fgr:contaminant}. In this case, as well, the concentration of species such as K$^+$, Na$^+$, and Ca$^+$ are amplified in the thin film region of the pattern sample. While the specific source introducing these species is unknown, these species, in particular K$^+$ and Na$^+$ which have unpaired electrons, can potentially facilitate electron tunneling at low temperatures.\cite{Muller_2019}
\section{Conclusions} \label{sec:conclusions}
Here we investigate the role that various fabrication procedures play in introducing dissipation mechanisms in quantum systems using time-of-flight secondary ion mass spectrometry (TOF-SIMS). In the case of an optimized deposition process for 2D transmon qubit (comparison to a previous generation of a 3D transmon qubit and a standard Nb RF cavitiy are provided in Figures S2 and S3), we find that following sputter deposition of Nb thin films, a thin layer of Si "floats" to the top surface, while alloyed Nb/Si regions are generated at the film/substrate interface. We also find that the lithography steps employed lead to the incorporation of impurity species including O$^-$, H$^-$, C$^-$, Cl$^-$, F$^-$, Na$^+$, Mg$^+$, and Ca$^+$ within the Nb thin film. As these species can appreciably alter the superconducting properties and/or introduce stochastic charge noise in these systems, extending coherence times will require future designs that employ new strategies to diminish these contaminants in the future.
\section*{Acknowledgements} \label{sec:acknowledgements}
This material is based upon work supported by the U.S. Department of Energy, Office of Science, National Quantum Information Science Research Centers, Superconducting Quantum Materials and Systems Center (SQMS) under the contract No. DE-AC02-07CH11359. This work made use of the EPIC facility of Northwestern University’s NU\textit{ANCE} Center, which received support from the Soft and Hybrid Nanotechnology Experimental (SHyNE) Resource (NSF ECCS-1542205); the MRSEC program (NSF DMR-1121262) at the Materials Research Center; the International Institute for Nanotechnology (IIN); the Keck Foundation; and the State of Illinois, through the IIN. This work made use of instruments in the Electron Microscopy Core of UIC’s Research Resources Center. The authors thank members of the Superconducting Quantum Materials and Systems (SQMS) Center for valuable discussion.
\bibliography{library}

% \begin{thebibliography}{4}
% \bibitem{Griffiths}
% D. J. Griffiths,
% \textit{Introduction to Electrodynamics}
% (Cambridge University Press, Cambridge, 2017).

% \bibitem{Fleming}
% A. Bobrinha,
% Revista Brasileira de Lorem Ipsum \textbf{23},
% 179 (2002).

% \bibitem{Feynman}
% R. P. Feynman, R. B. Leighton and M. Sands,
% \textit{Lições de Física de Feynman}
% (Editora Bookman, Porto Alegre, 2008).

% \bibitem{Jackson-CE}
% J. D. Jackson,
% \textit{Classical Electrodynamics}
% (John Wiley \& Sons, Danvers, 1999).
% \end{thebibliography}

\appendix*
\section{STEM Imaging Conditions} \label{sec:appendix}
Electron-transparent lamella of Nb thin films deposited on Si wafers were fabricated using a focused ion beam. These samples were imaged using a JEOL ARM 200CF ARM S/TEM operated at 200 kV. The camera length is set to 8 cm and the condenser aperture is selected to provide a convergence semi-angle of 30 mrad with beam current ~20 pA in order to minimize beam-induced damage. 

\setcounter{figure}{0}
\makeatletter 
\renewcommand{\thefigure}{S\@arabic\c@figure}
\makeatother

\begin{figure*}
\includegraphics[width=7in]{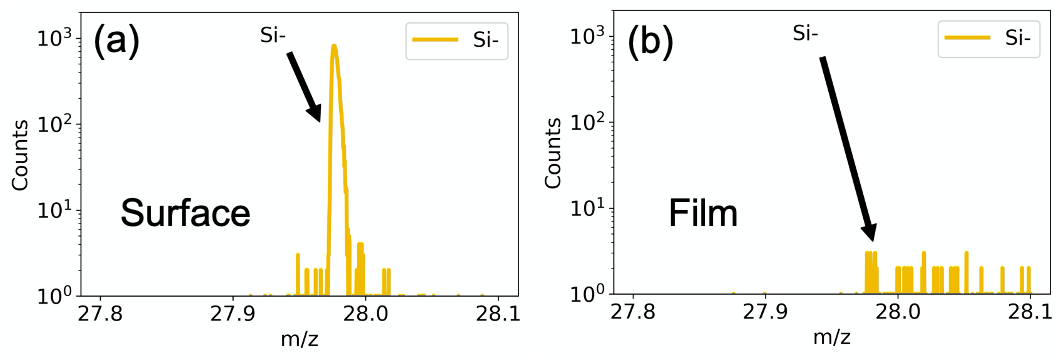}
\caption{A mass spectra comparison taken from (a) surface and (b) Nb film. Signal from Si is greatly attenuated following the first few nanometers.}
\label{fgr:Si}
\end{figure*}

\begin{figure*}
\includegraphics[width=7in]{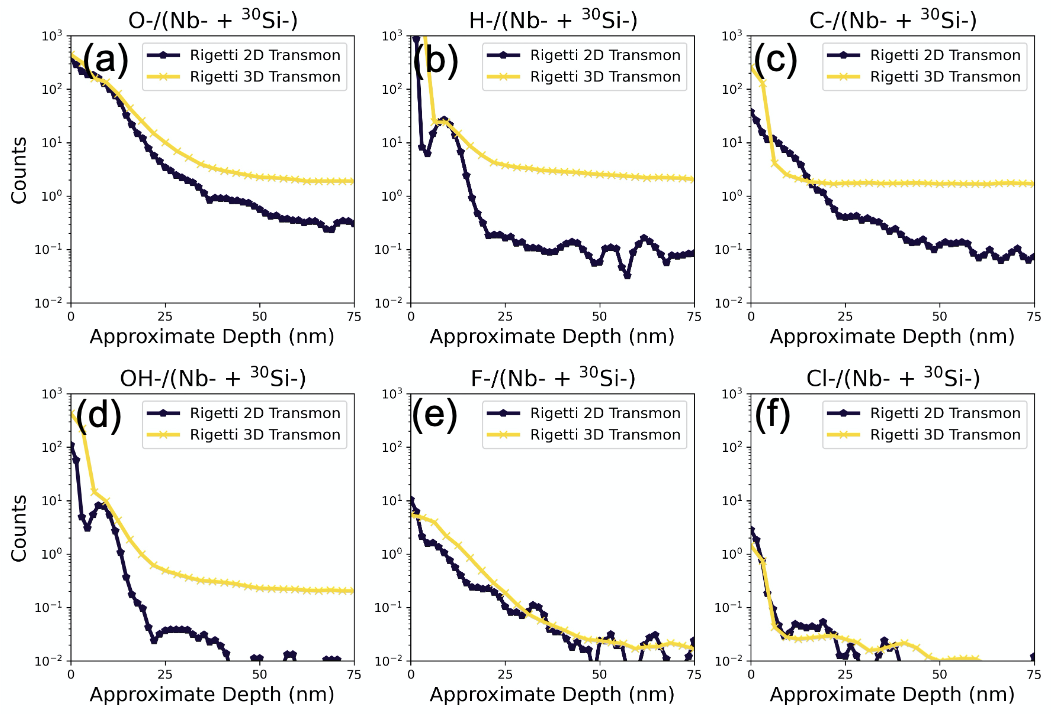}
\caption{A depth profile comparison of the 2D and 3D transmon structures fabricated using different fabrication procedures. 2D transmon architectures were fabricated with optimized baking and deposition conditions to minimize concentration of oxygen and hydrocarbon species.(a) O-, (b) H-, (c) C-, (d) OH-, (e) F-, and (f) Cl- signals are plotted as a function of depth for the two samples investigated.}
\label{fgr:2D_3D}
\end{figure*}

\begin{figure*}
\includegraphics[width=7in]{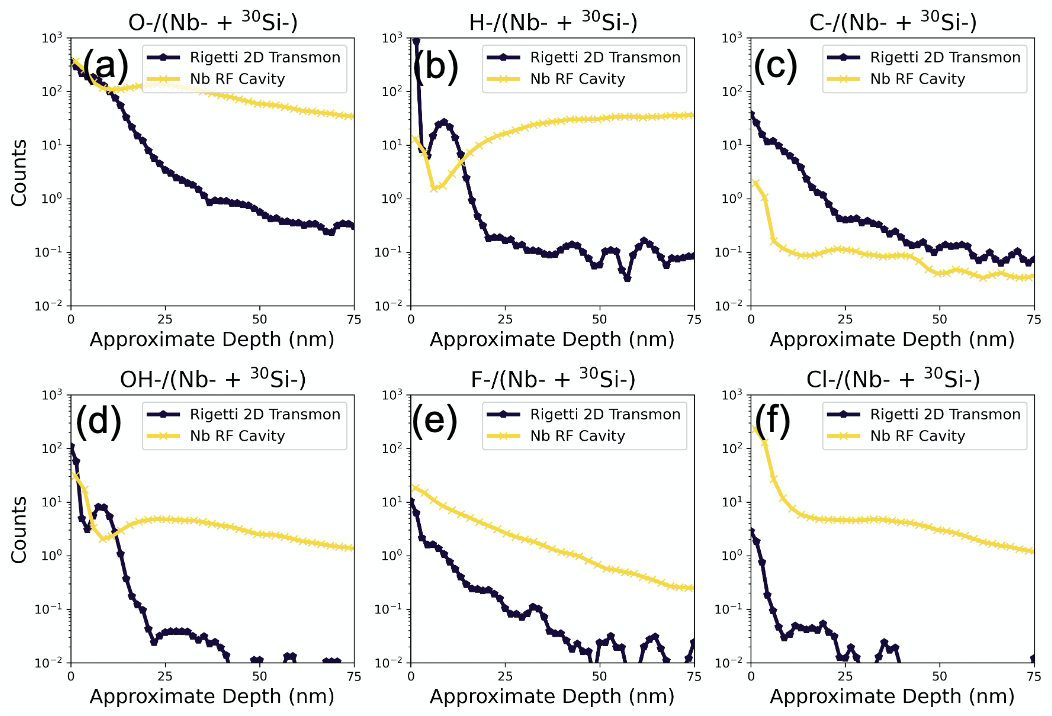}
\caption{A depth profile comparison of the 2D transmon structure and a standard Nb cavity.(a) O-, (b) H-, (c) C-, (d) OH-, (e) F-, and (f) Cl- signals are plotted as a function of depth for the two samples investigated.}
\label{fgr:cavity}
\end{figure*}

\end{document}